\documentclass[showpacs,aps,prl,superscriptaddress,letterpaper]{revtex4}
\usepackage{graphicx}
\usepackage{dcolumn}
\usepackage{amsmath}
\usepackage{epsfig}
\long\def\inst#1{\par\nobreak\kern 4pt\nobreak
    {\itshape #1}\par\vskip 10pt plus 3pt minus 3pt}
\RequirePackage{xspace}

\def\babar{\mbox{\slshape B\kern-0.1em{\smaller A}\kern-0.1em
    B\kern-0.1em{\smaller A\kern-0.2em R}}}
\def\Kbar    {\kern 0.18em\overline{\kern -0.18em K}{}\xspace}

\def\Kz      {\ensuremath{K^0}\xspace}
\def\Kzb     {\ensuremath{\Kbar^0}\xspace}
\def\KzKzb   {\ensuremath{\Kz {\kern -0.16em \Kzb}}\xspace}

\def\Ks     {\ensuremath{K_S}\xspace}
\def\Kl     {\ensuremath{K_L}\xspace}
\def\KsKs   {\ensuremath{\Ks {\kern -0.16em \Ks}}\xspace}
\def\KlKl   {\ensuremath{\Kl {\kern -0.16em \Kl}}\xspace}
\def\KsKl   {\ensuremath{\Ks {\kern -0.16em \Kl}}\xspace}
\def\KlKs   {\ensuremath{\Kl {\kern -0.16em \Ks}}\xspace}
\def\Dbar    {\kern 0.18em\overline{\kern -0.18em D}{}\xspace}
\def\Dz      {\ensuremath{D^0}\xspace}
\def\Dzb     {\ensuremath{\Dbar^0}\xspace}
\def\DzDzb   {\ensuremath{\Dz {\kern -0.16em \Dzb}}\xspace}

\newcommand{\DsP}{\ensuremath{D_s^+}\xspace}
\newcommand{\DsM}{\ensuremath{D_s^-}\xspace}
\newcommand{\DspDsm}{\ensuremath{\DsP {\kern -0.16em \DsM}}\xspace}
\newcommand{\Dp}{\ensuremath{D^+}\xspace}
\newcommand{\Dm}{\ensuremath{D^-}\xspace}
\newcommand{\DpDm}{\ensuremath{\Dp {\kern -0.16em \Dm}}\xspace}

\def\Bbar    {\kern 0.18em\overline{\kern -0.18em B}{}\xspace}

\def\Bz      {\ensuremath{B^0}\xspace}
\def\Bzb     {\ensuremath{\Bbar^0}\xspace}
\def\BzBzb   {\ensuremath{\Bz {\kern -0.16em \Bzb}}\xspace}
\def\Bu      {\ensuremath{B^+}\xspace}
\def\Bub     {\ensuremath{B^-}\xspace}

\def\BpBm    {\ensuremath{\Bu {\kern -0.16em \Bub}}\xspace}

\def\Dp      {\ensuremath{D^+}\xspace}

\newcommand{\optbar}[1]{\shortstack{{\tiny (\rule[.4ex]{1em}{.1mm})}
  \\ [-.7ex] $#1$}}
\def\BorBbar    {\kern 0.18em\optbar{\kern -0.18em B}{}\xspace}
\def\DorDbar    {\kern 0.18em\optbar{\kern -0.18em D}{}\xspace}
\def\KorKbar    {\kern 0.18em\optbar{\kern -0.18em K}{}\xspace}

\def\pep2{PEP-II}
\mathchardef\Upsilon="7107
\def\Y#1S{\ensuremath{\Upsilon{(#1S)}}\xspace}

\begin{document}

\title{\large \bfseries \boldmath Probing dark force at BES-III/BEPCII }
\author{Hai-Bo Li}\email{lihb@ihep.ac.cn}
\author{Tao Luo}\email{luot@ihep.ac.cn}
\affiliation{Institute of High Energy Physics, P.O.Box 918,
Beijing  100049, China}


\date{\today}


\begin{abstract}
We study an experimental search of a GeV scale vector boson at
BES-III/BEPCII. It is responsible for mediating a new U(1)$_d$
interaction, as recently exploited in the context of weakly
interacting massive particle dark matter. At low energy $e^+ e^-$
colliders this dark state can be efficiently probed. We discuss
the direct productions of this light vector $U$ boson and the
decay of this state with BES-III data and its foreseen larger
data. In particular, we show that Higgs'-strahlung in the dark
sector can lead to multilepton signatures, which probe the physics
range for kinetic mixing parameter $\epsilon \sim 10^{-4}
-10^{-3}$ over a large portion of the parameter space.

\end{abstract}


 \maketitle

\section{Introduction}

 In recent years, many astrophysical and cosmological experiments
 have indicated the presence of dark matter (DM). The leading
 candidate for this DM is a particle~\cite{darko}. There are many
 pieces of evidence that the DM component of the universe may be
 light and at GeV scale. The DM of this scale
has been invoked to explain several recent experimental
results~\cite{dama1,int1,int2,int3}, including the annual
modulation of the DAMA/LIBRA~\cite{dama} and the 511 keV photon
signal from the SPI/INTEGRAL~\cite{int}. Sub-GeV scale bosons have
also been used to interpret the excess in the cosmic ray positron
reported by PAMELA~\cite{pamela} and the total electron and
positron flux measured by ATIC~\cite{atic}, as well as preliminary
result from Fermi-LAT~\cite{fermi}.

The light U boson may couple to the SM charged particles with a
much suppressed coupling which has been considered in various
contexts~\cite{int1,uboson-model1:gninenko:2001,uboson-model2:fayet:2004,uboson-model2:dorokhov:2007,uboson-model3:zhush:2007,uboson-model4:drees:2006}.
We consider the new Abelian gauge group $U(1)_d$ which has a
gauge-invariant kinetic mixing with the SM hypercharge
$U(1)_Y$~\cite{nima:2008,holdom:1986,dienes:1997}. After
electroweak symmetry breaking, we have the Lagrangian
\begin{equation}
{\cal L} = {\cal L}_{SM} + \epsilon_Y F^{Y,\mu\nu}F^d_{\mu\nu} +
m^2_{U} A^{d, \mu}A^d_{\mu},
 \label{mixingl}
\end{equation}
where ${\cal L}_{SM}$ is the SM Lagrangian, $F^Y_{\mu\nu}$ and
$F^d_{\mu\nu}$ are the field strength for $U$ boson and the gauge
boson $B$ of U(1)$_Y$, $A^d$ is the gauge field of a massive dark
$U(1)_d$ gauge group~\cite{holdom:1986}. The second term in
Eq.(\ref{mixingl}) is kinetic mixing operator, and $\epsilon \sim
10^{-8} - 10^{-2}$ is generated at any scale by loops of heavy
fields charged under both $U(1)$s. In a supersymmetry theory, the
kinetic mixing operator induces a mixing between the D-terms
associated with $U(1)_d$ and $U(1)_Y$. The hypercharge D-term gets
a vacuum expectation value from electroweak symmetry breaking and
induces a weak-scale effective Fayet-Iliopoulos term for $U(1)_d$.
Consequently, the $U(1)_d$ symmetry breaking scale is naturally
suppressed by loop factor or by $\sqrt{\epsilon}$, leading to MeV
to GeV-scale U boson mass~\cite{nima:2008,dienes:1997}. The
parameters of concern in this Letter are $\epsilon$ and $m_{U}$.
It is also natural to conceive the existence of an elementary
Higgs-like boson, the $h^\prime$, which spontaneously breaks the
symmetry as argued in reference~\cite{batell:2009}. Although the
$U$ boson will in general have a substantial branching ratio to
lepton pairs, the decays of the Higgs' will depend on its mass
relative to that of the $U$ boson. If the Higgs' is light it will
decay via loop processes to leptons and possible hadrons, in which
case it is long-lived and will most likely appear as missing
energy. Otherwise, if it is heavy, it will decay to double $U$
bosons, and finally can be seen in multiple lepton final states.

An interesting consequence of the above hypotheses is that they
must induce observable effects in low energy $e^+ e^-$ colliders,
such as the existing or future super-flavor factories. A more
comprehensive discussion of these effects can be found in
references ~\cite{batell:2009},~\cite{essig:2009},
\cite{wang:2009} and \cite{Bjorken:2009mm}.

Both the $U$ boson and the $h^\prime$ can be produced at BEPCII,
if their masses are less than charmonium states. Although the
luminosity of BEPCII is order of magnitude lower than that at B
factories, the production cross-sections scale as $1/s$, which
essentially compensates the lower luminosity at BEPCII.
Furthermore, BEPCII is better suited for detecting particles with
mass around 1 GeV, and also is a great lab to look for $U$ boson
and $h^\prime$ in $\psi$ decays, which will be produced with huge
sample ($10^{10}$ $J/\psi$ decays events per year at BEPCII).
These considerations motivate searching for dark-sector events in
various channels at BEPCII/BES-III. We will mainly argue the
following processes at BES-III:
\begin{itemize}
\item $e^+ e^- \rightarrow \gamma + U \rightarrow \gamma l^+l^-$,
where $U \rightarrow l^+l^-$, $l$ could be electron or muon; \item
$J/\psi \rightarrow U l^+l^- \rightarrow 4l$; \item $\psi(2S)
\rightarrow U \chi_{c1,2} \rightarrow e^+e^-\chi_{c1,2}$; \item
$J/\psi \rightarrow U h^\prime \rightarrow l^+l^- +$missing
energy; or $J/\psi \rightarrow U h^\prime \rightarrow 3 U
\rightarrow 6 l$;
\end{itemize}

\section{Status of BES-III}

The BES-III detector consists mainly of a cylindrical main draft
chamber (MDC) with momentum resolution $\sigma_{p_t}/p_t \sim
0.5\%$ for a charged particle with momentum at 1.0 GeV, the
time-of-flight (TOF) system with two layers of plastic
scintillator counters located outside of MDC, and highly hermetic
electromagnetic calorimeter (EMC) with energy resolution of
$\sigma_E /E = 2.5\%/\sqrt{E(\mbox{GeV})}$~\cite{besiiidetector}.
The MDC has its first sensitive layer at a radius of 6.0 cm from
the interaction point (IP), and the MDC combined with a B field of
1.0 T, provide a precise momentum measurements of charged
particles with transverse momentum greater than 50 MeV. Photons of
energy down to 20 MeV and polar angle in the range
$21^0<\theta<159^0$ can be detected with good efficiency by EMC.

BES-III has so far acquired about $1.0\times 10^{8}$ and
$2.0\times 10^{8}$ at $\psi(2S)$ and J/$\psi$ peaks, respectively.
In the next few years, by year 2014, $10^{10}$ events on J/$\psi$
peak and $3\times 10^{9}$ events on $\psi(2S)$ peak will be
collected. The expected data samples per year running at BES-III
are summarized in table~\ref{tab:1}. For this Letter, the
sensitivity studies are based on 3 fb$^{-1}$ luminosity at the
$J/\psi$ or $\psi(2S)$ peak and 20fb$^{-1}$ at the $\psi(3770)$
peak for the searching for the dark sector.
\begin{table}[htbp]
\caption{$\tau$-Charm productions at BEPCII in one year's running
($10^7s$).} \label{tab:1}
\begin{tabular}{@{}lll}
\hline
               & Central-of-Mass    & \#Events  \\
Data Sample    & (MeV)                   & per year  \\
\hline
$J/\psi$ &  3097     & $10\times 10^9$\\
$\tau^+\tau^-$   & 3670  & $12\times 10^6$ \\
$\psi(2S)$ & 3686  & $3.0\times 10^9$ \\
$\DzDzb$ & 3770  & $18\times 10^6$ \\
$\DpDm$ & 3770  & $14\times 10^6$ \\
$\DspDsm$ & 4030  & $1.0\times 10^6$ \\
$\DspDsm$ & 4170  & $2.0\times 10^6$ \\
\hline
\end{tabular}
\end{table}

\section{Reach of $U$ boson search at BES-III}
\label{sec:production}

In this section, we discuss the constraints and discovery
potential for the $U$ boson at Beijing electron-positron collider
II (BEPCII)~\cite{besiiidetector}. Since the $U$ boson couples
mainly to the SM electromagnetic current~\cite{wang:2009}. Its
production at BEPCII is the same as that of photon, although with
a much suppressed rate. Therefore, any process which produces a
large number of detectable photons will have a chance to produce
$U$ boson as well.
\\
\vspace*{0.2cm}

{\bf 1. $e^+e^- \rightarrow U \gamma$ events }\\
\vspace*{0.2cm}

The process of $e^+e^- \rightarrow U\gamma$ is one of the most
interesting processes, which had been discussed previously in
references~\cite{uboson-model2:fayet:2004,uboson-model3:zhush:2007,uboson-model4:drees:2006,wang:2009}.
It has the advantage of being independent of details of the Higgs'
sector~\cite{batell:2009}. The on-shell $U$ boson will decay to a
pair of leptons, leading to a signal of $l^+l^- \gamma$. The SM
background $e^+e^-\rightarrow \gamma^* \gamma \rightarrow
l^+l^-\gamma$, although large for this process, is not a severe
problem as the kinematics of the signal are quite distinct. The
invariant mass of the lepton pair is just within a single bin due
to the tiny width of the vector and can be distinguished from the
QED background.
\begin{figure}
 \epsfig{file=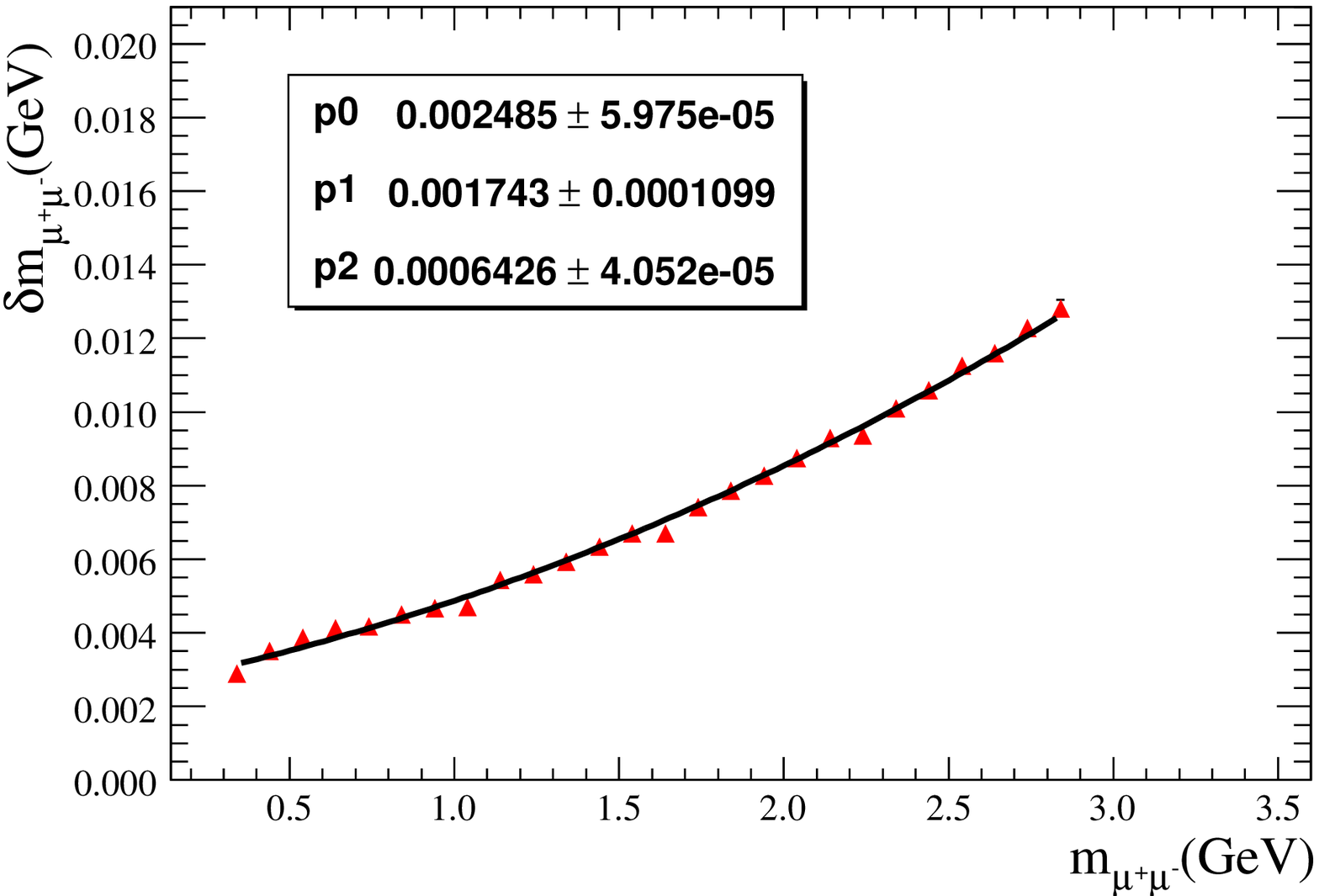,width=8cm,height=5cm}
\epsfig{file=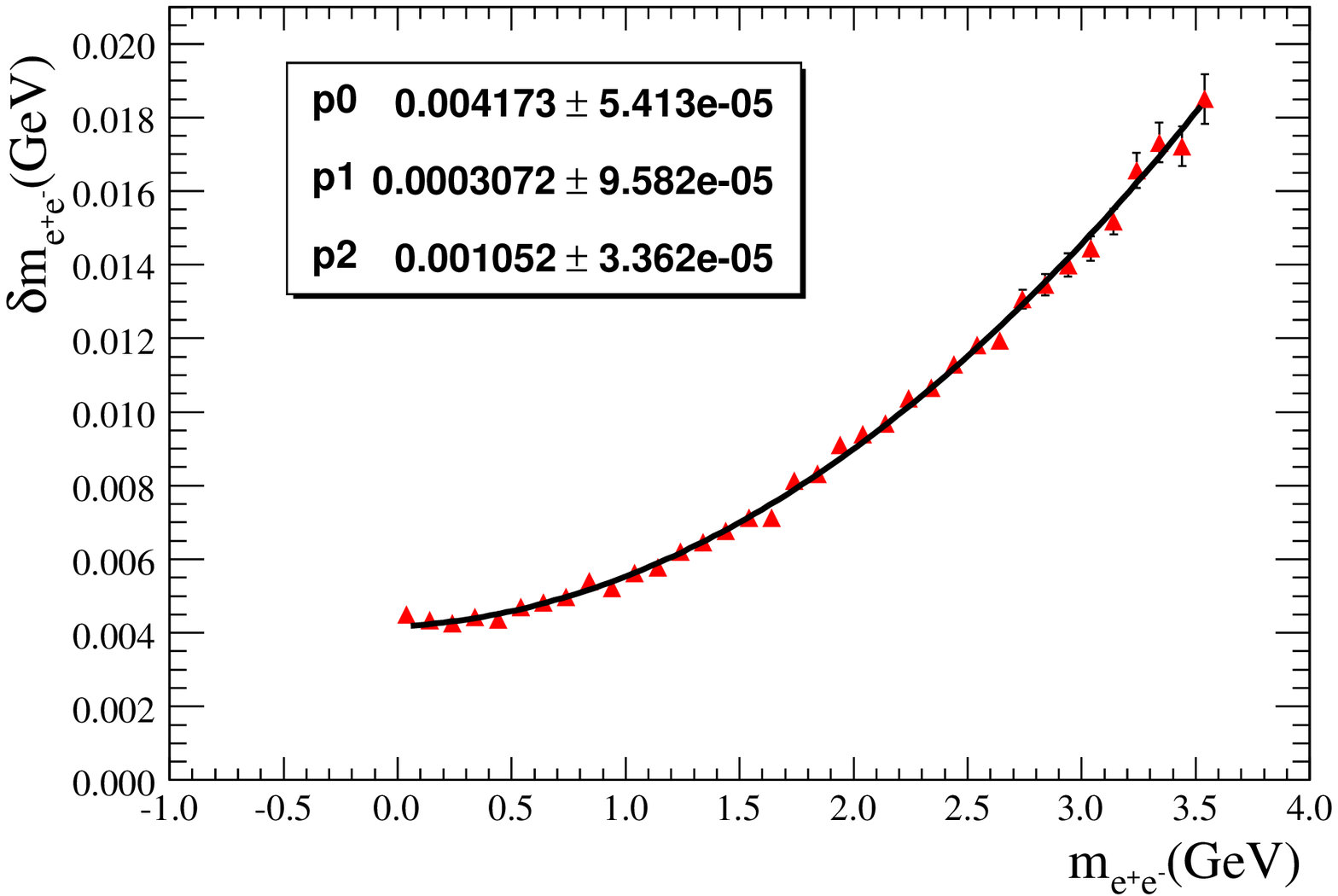,width=8cm,height=5cm}
\caption{Resolution in invariant mass of lepton pairs at BES-III,
as a function of $m_{\mu^+\mu^-}$(left) and $m_{e^+e^-}$ (right).}
\label{fig:0}
\end{figure}
Over a large range of parameters, the cross-sections for the
production of $U$ boson scale as $\sigma_s = \alpha^2 \epsilon^2/s
\sim \sigma_0 \epsilon^2$, where $\sigma_s$ is the signal
cross-section for the $e^+e^- \rightarrow U \gamma$, and
$\sigma_0$ is the cross-section for the analogous QED process
$e^+e^- \rightarrow \gamma \gamma$. The value of the cross-section
for $e^+e^- \rightarrow \gamma \gamma$ is about 42 nb at
$\sqrt{s}=3.773$ GeV, and assuming $\epsilon = 10^{-3}$, the
cross-section for $U$ boson production could be 42 fb.  We
estimate the expected numerical results based on the significance
\begin{eqnarray}
\frac{S}{\sqrt{B}} = \frac{\sigma_s\times {\cal
L}}{\sqrt{\sigma_{ith-bin}  \times {\cal L}}} \times
BR(U\rightarrow l^+l^-) = \sqrt{{\cal L}}\frac{\epsilon^2
\sigma_0}{\sqrt{\sigma_{ith-bin}}}\times BR(U\rightarrow l^+l^-),
\label{cross:gen}
\end{eqnarray}
where ${\cal L}$ is integrated luminosity at BES-III, the
$BR(U\rightarrow l^+l^-)$ is the branching ratio of $U$ boson
decay into lepton pair which is discussed in
reference~\cite{batell:2009}, and $\sigma_{ith-bin}$ is the
effective cross-section from QED background $e^+e^- \rightarrow
\gamma^* \gamma \rightarrow e^+e^- \gamma$ in the $i$-th bin on
the distribution of $m_{l^+l^-}$. The effective cross-section can
be estimated based on the resolution function at BES-III, we have
\begin{eqnarray}
\sigma_{ith-bin} = \sigma(e^+e^- \rightarrow \gamma^* \gamma
\rightarrow e^+e^- \gamma) \frac{N_{ith-bin}}{N_{total}},
\label{eff:cross}
\end{eqnarray}
where $N_{ith-bin}$ is the number of events in the $i$-th bin with
size $\delta m$ and $N_{total}$ is the total number of observed
events of QED background. The size of the bin is in the window of
the size $\delta m$ around $m_{l^+l^-} = m_U$, here $\delta m$ is
the mass resolution of $m_{l^+l^-}$ in the BES-III detector. The
width of $U$ boson is much smaller than the typical detector
resolution, so it is important to understand the mass resolution
of BES-III. We simulated the QED background $e^+e^- \rightarrow
e^+e^- \gamma$ by considering full BES-III detector within GEANT4
framework. We obtain the following resolution functions:
\begin{eqnarray}
\delta m (\mu^+\mu^-) =\left( 2.5 + 1.7 \left(\frac{m_U}{1.0
\mbox{GeV}}\right)+0.6\left(\frac{m_U}{1.0 \mbox{GeV}}\right)^2
\right)\,\, (\mbox{MeV}), \label{reso:leptonmumu}
\end{eqnarray}
\begin{eqnarray}
\delta m (e^+e^-) =\left( 4.1 + 0.3 \left(\frac{m_U}{1.0
\mbox{GeV}}\right)+1.1\left(\frac{m_U}{1.0 \mbox{GeV}}\right)^2
\right)\,\, (\mbox{MeV}). \label{reso:leptonee}
\end{eqnarray}
To obtain the resolution function, we require that the
reconstructed tracks must be within the the fiducial volume of the
MDC, and the photon must be in the EMC.  In order to reconstruct
charged tracks with good quality and neutral tracks without
dilution of detector noises, we ask $p_t
> 80 \mbox{MeV}$ and $E_{\gamma} > 20 \mbox{MeV}$, where $p_t$ and $E_{\gamma}$
are transverse momentum of charged track in the MDC and deposit
energy of photon in the EMC, respectively. We assume that the
inefficiency along the invariant mass distribution of $m_{l^+l^+}$
is flat, and the resolution, $\delta m$ is plotted against
$m_{l^+l^+}$ in Fig.~\ref{fig:0}.

\begin{figure}
 \epsfig{file=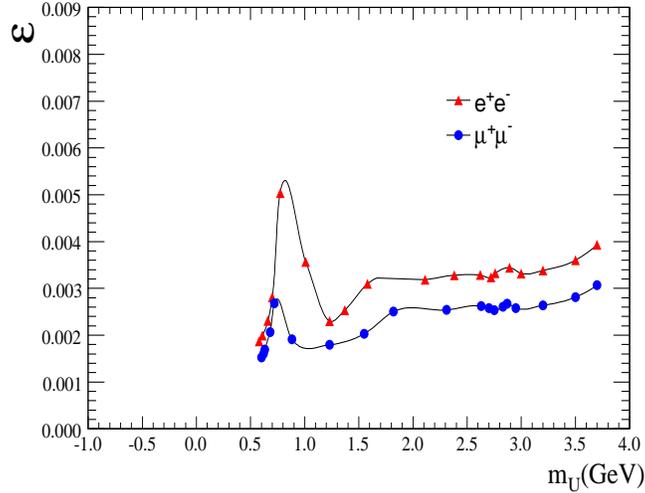,width=9cm,height=7cm}
\caption{Illustrative plot of the reach of vector boson at BES-III
in the channel of $e^+e^- \rightarrow \gamma U \rightarrow \gamma
l^+l^-$. Note that the sensitivity of $\epsilon$ is becoming worse
at large mass of U boson due to the large background from Bhabha
scattering.  } \label{fig:1}
\end{figure}

In Fig.~\ref{fig:1}, we show the reach of the parameter $\epsilon$
by defining $S/\sqrt{B} = 5$ in Eq.~(\ref{cross:gen}). The signal
events are generated with different choice of $m_U$, and passing
the same cuts as in the above discussion. We count the number of
events in the window of one resolution $\delta m$ for both signal
and background. It is important to know the $BR(U\rightarrow
e^+e^-)$ and $BR(U\rightarrow \mu^+\mu^-)$ in the estimation of
the reach in Fig.~\ref{fig:1}. The branching ratios of $U$ boson
decay into lepton pair and hadron states, such as $\pi^+\pi^-$
have been discussed in reference~\cite{batell:2009}, in which the
hadronic decay rate is extracted from the R values measured in
$e^+e^- \rightarrow$hadron processes~\cite{batell:2009}. The reach
of $l^+l^-$ is degraded if the $U$ boson mass is in the region of
$\rho$ resonance, because the branching ratio of $U\rightarrow
l^+l^-$ is becoming small due the open of hadron $\pi^+\pi^-$
channel in that region. For this estimation, we assume 20fb$^{-1}$
data sample collected at $\psi(3770)$ peak at BES-III.

There is another background which is not considered here for
$e^+e^- \rightarrow U \gamma$. It is a relevant instrumental
background that has to be taken into account for this channel,
namely $e^+e^- \rightarrow \gamma \gamma$ with subsequent
conversion of one of the two photons on the beam pipe and inner
wall of the MDC, with a probability of $\sim$2\%. Since, for the
$\gamma$-conversion, the $e^+e^-$ is not from interaction point,
the invariant mass is peak near zero and the open angle between
two electrons is small, this background can be rejected by
requiring cuts on the reconstructed invariant mass and vertex of
the pair as done for instance in the analysis of
\cite{chi:c:bes2:2006}. For $m_U$ larger than 500 MeV, assuming
the energy resolution in the calorimeter is $\sim$20 MeV for
photons of 500 MeV, one finds that this background is negligible.
In Fig.~\ref{fig:1}, we only plot the reach for the invariant mass
larger than 500 MeV.

\vspace*{0.3cm}

{\bf 2. $U$ boson in $J/\psi$ decays  }\\
\vspace*{0.2cm}

\begin{figure}
 \epsfig{file=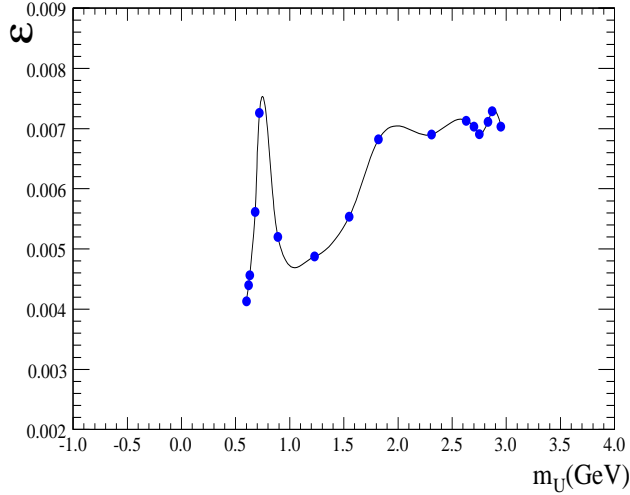,width=9cm,height=7cm}
\caption{Illustrative plot of the reach of vector boson at BES-III
in the channel of $J/\psi \rightarrow e^+e^- U \rightarrow e^+e^-
\mu^+\mu^-$.  } \label{fig:2}
\end{figure}

A study of $U$ boson in the decay of $J/\psi\rightarrow e^+e^- U$
for in a lower mass range, $m_U \leq 50 $MeV has been considered
in Ref.~\cite{uboson-model3:zhush:2007}. We will have $10^{10}$
$J/\psi$ events at BES-III with one year
luminosity~\cite{besiiidetector}. The background process $J/\psi
\rightarrow \gamma e^+e^-$ has branching ratio of $8.4\times
10^{-3}$ which is measured by Fermilab E760 experiment with
$E_{\gamma}
> 100$ MeV~\cite{fermi:1996}, which is consistent with the SM
predictions. This decay rate is large enough and can be used to
detect the $U$ boson via $J/\psi \rightarrow e^+e^- U$ channel.
The mode $J/\psi \rightarrow \gamma \mu^+ \mu^-$ has also been
measured~\cite{brown:1983}. With $10^{10}$ $J/\psi$ events at
BES-III, the reach in the channel $J/\psi \rightarrow e^+e^- U$
could be at the level of $\epsilon \sim 10^{-3}$, which is
competitive with the process $e^+e^- \rightarrow U \gamma$. Of
course there could be also an $e^+e^- l^+l^-$ final state, with a
large background from QCD process $e^+e^- \rightarrow e^+e^-
l^+l^-$
\begin{eqnarray}
\sigma_{e^+e^-\rightarrow e^+e^- l^+l^-} = \frac{\alpha^4}{\pi
m_l^2}\left( \mbox{log}\frac{s}{m^2_e}\right)^2
\mbox{log}\frac{s}{m^2_l};\,\, \,\, l= e \,\,\mbox{or} \,\,\mu,
\label{ee:leptonmumu}
\end{eqnarray}
for the $e^+e^-\rightarrow e^+e^- \mu^+\mu^-$ process, the
production cross-section is about 69 nb at $J/\psi$ peak. However,
the electrons in the two-photon process move primarily in the
direction of the beam pipe. Small angle and invariant mass cuts
should help to reduce the background. Moreover, the signal $U$
boson decay events will have a significant peak, and make this
kind of background rejection somewhat easier. In Fig.~\ref{fig:2},
we show the reach of the parameter $\epsilon$ by defining
$S/\sqrt{B} = 5$ in the channel $J/\psi \rightarrow e^+e^- U$ ($U
\rightarrow \mu^+\mu^-$), in which backgrounds from $J/\psi
\rightarrow \gamma^* e^+e^- \rightarrow \mu^+\mu^- e^+e^-$ and QED
precess $e^+e^- \rightarrow e^+e^- \mu^+\mu^-$ are considered.
Again, the reach is degraded if the $U$ boson mass is in the
region of $\rho$ resonance~\cite{batell:2009}.

 There are another interesting channels such as $J/\psi
 \rightarrow \eta U$ and $\eta^\prime U$ which can be used to search for $U$ boson, since the branching ratios
 of $J/\psi$ decay into $\eta + X$ and $\eta^\prime +X$ are
 percentage level. The reach of these channels could be at the level
 of $\epsilon \sim 10^{-3}$ at BES-III.  The irreducible
 backgrounds in these channels are from $J/\psi \rightarrow
 \eta/\eta^\prime V$, where the $V$ denotes $\rho^0$, $\phi$ and
 $\omega$ vector mesons which can decay into $e^+e^-$ pair.

\vspace*{0.3cm}

{\bf 3. $U$ boson in $\psi(2S) \rightarrow U +\chi_{c1,2}$ decays  }\\
\vspace*{0.2cm}

\begin{figure}
 \epsfig{file=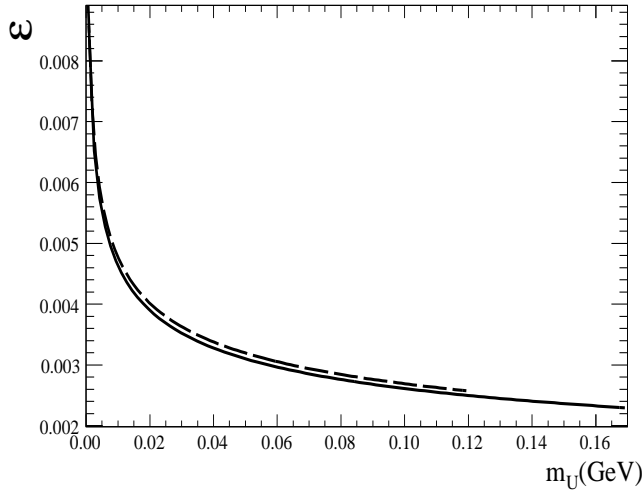,width=9cm,height=7cm}
\caption{Illustrative plot of the reach of vector boson at BES-III
in the channel of $\psi(2S) \rightarrow U \chi_{c1}$ (solid curve)
and $\psi(2S) \rightarrow U \chi_{c2}$ (dashed curve), where U
decay into $e^+e^-$. } \label{fig:3}
\end{figure}

At the BES-III experiment, we will collect about $3\times 10^{9}$
$\psi(2S)$ events in one year luminosity as listed in
table~\ref{tab:1}. Two interesting decay modes are $\psi(2s)
\rightarrow \gamma \chi_{c1}$ and $\gamma \chi_{c2}$ with
branching fractions at the level of 10\%. It is important to note
that both $\chi_{c1}$ and $\chi_{c2}$ can be reconstructed from
$\chi_{c1,2} \rightarrow \gamma J/\psi \rightarrow \gamma l^+l^-$,
and the combined branching fraction could be about 4.0\%. The
narrow $J/\psi$ peak can be used to tag $\chi_{c1,2}$ decays.
There will be huge amount of events observed in $\psi(2S)$ decays.
There is about 176 MeV (130 MeV) of phase space for $\psi(2S)
\rightarrow \gamma \chi_{c1}$ ($\psi(2S) \rightarrow \gamma
\chi_{c2}$). It will be interesting to look for a low mass,
$\sim$100 MeV, $U$ boson in these modes.

The dominant background is $\psi(2S) \rightarrow \gamma^*
\chi_{c1,2} \rightarrow e^+e^- \chi_{c1,2}$, where $m_{e^+e^-} =
q_{\gamma^*} = m_U$. Since the $BR(\psi(2S) \rightarrow \gamma
\chi_{c1,2})$ is well measured, typically, we have
\begin{eqnarray}
BR(\psi(2S) \rightarrow \gamma^* \chi_{c1,2} \rightarrow e^+e^-
\chi_{c1,2}) \sim 10^{-2} \times BR(\psi(2S) \rightarrow \gamma
\chi_{c1,2}).
 \label{psi2s:brback}
\end{eqnarray}
As discussed in Ref.~\cite{wang:2009}, the number of background
events in the window of $\delta m$ (resolution of $m_{e^+e^-}$)
around $m_{e^+e^-} = m_U$ is about
\begin{eqnarray}
N_{B} = N_{\psi(2S)} BR(\psi(2S)\rightarrow \gamma^*
\chi_{c1,2}\rightarrow e^+e^-\chi_{c1,2}) \frac{\delta m}{m_U}
\frac{1}{log[(m_{\psi(2S)}-m_{\chi_{c1,2}})/2m_e]},
 \label{psi2s:back}
\end{eqnarray}
where $\delta m$ is the resolution for reconstructed invariant
mass $m_{e^+e^-}$ at the BES-III experiment; $m_{\psi(2S)}$,
$m_{\chi_{c1,2}}$ and $m_e$ are the nominal masses of $\psi(2S)$,
$\chi_{c1,2}$ and electron, respectively; and $N_{\psi(2S)}$ is
the total number of $\psi(2S)$ decay events.  By replacing the
photon by $U$ boson, the signal rate can be estimated to be
$BR(\psi(2S) \rightarrow U \chi_{c1,2}) \sim \epsilon^2
BR(\psi(2S) \rightarrow \gamma \chi_{c1,2})$. Thus, the expected
number of signal events is about
\begin{eqnarray}
N_{S} = N_{\psi(2S)}\times \epsilon^2 BR(\psi(2S)\rightarrow
\gamma \chi_{c1,2})\times BR(U\rightarrow e^+e^-),
 \label{psi2s:sig}
\end{eqnarray}
where $BR(U\rightarrow e^+e^-)=1$ since the $U$ boson can only
decay into electron-positron pair due to the limit of phase space.
Combining Eqs.~(\ref{psi2s:back}) and (\ref{psi2s:sig}), we obtain
the signal significance as
\begin{eqnarray}
\frac{S}{\sqrt{B}} &=& \frac{N_S}{\sqrt{N_B}} \\ \nonumber
&=&\sqrt{N_{\psi(2S)}}\frac{\epsilon^2 BR(\psi(2S)\rightarrow
\gamma \chi_{c1,2})BR(U\rightarrow
e^+e^-)}{\sqrt{BR(\psi(2S)\rightarrow \gamma^*
\chi_{c1,2}\rightarrow e^+e^-\chi_{c1,2})}}\times
\sqrt{\frac{m_U}{\delta
m}log\frac{(m_{\psi(2S)}-m_{\chi_{c1,2}})}{2m_{e}}}.
 \label{psi2s:uchic}
\end{eqnarray}
Therefore, with $3\times 10^{9}$ $\psi(2S)$ events, the reach for
$U$ boson searching can be $\epsilon \sim 10^{-3}$ in $\psi(2S)
\rightarrow U \chi_{c1,2}$ decay.  In Fig.~\ref{fig:3}, we show
the reach of the parameter $\epsilon$ by defining $S/\sqrt{B} = 5$
for different mass of $U$ boson.

From experimental side of view, one can measure the following
ratio:
\begin{eqnarray}
{\cal R} = \frac{BR(\psi(2S) \rightarrow
U\chi_{c1,2})}{BR(\psi(2S)\rightarrow \gamma \chi_{c1,2}) },
 \label{psi2s:r}
\end{eqnarray}
where the $\chi_{c1,2}$ can be reconstructed in the $\chi_{c1,2}
\rightarrow \gamma J/\psi \rightarrow \gamma l^+l^-$ decay. Since
the width of $J/\psi$ is narrow, we use a kinematic fit that
constrains the decay of $J/\psi \rightarrow l^+l^-$ to $J/\psi$
mass, so that the resolution of invariant mass $m_{\gamma l^+l^-}$
can be greatly improved.  At BEPCII, the central-of-mass energy is
known at $\psi(2S)$ peak, one can define the missing mass squared,
$\mbox{MM}^2$, recoiling against the $\chi_{c1,2}$ tagged by
$\gamma J/\psi$
\begin{eqnarray}
\mbox{MM}^2 = (m_{\psi(2S)} - E_{\chi_{c1,2}})^2 - (- {\bf
p}_{\chi_{c1,2}})^2, \label{psi2s:r}
\end{eqnarray}
where $E_{\chi_{c1,2}}$ and ${\bf p}_{\chi_{c1,2}}$ are the energy
and three-momentum of the fully reconstructed $\chi_{c1,2}$. Real
$\psi(2S) \rightarrow \gamma \chi_{c1,2}$ will congregate near
zero $\mbox{MM}^2$, while real $\psi(2S) \rightarrow U
\chi_{c1,2}$ will peak near $\mbox{MM}^2=m^2_U$ which deviated
from zero. By looking at the missing mass squared $\mbox{MM}^2$,
one can extract the ratio ${\cal R}$. In this method, the $U$
boson decay and radiative photon are missed in the reconstruction.
We note that the reconstruction efficiencies are the same for both
decays of $\psi(2S) \rightarrow U \chi_{c1,2}$ and $\psi(2S)
\rightarrow \gamma \chi_{c1,2}$, and the uncertainties due to the
reconstructions of charged tracks and neutral tracks cancel in the
measurement of the ratio ${\cal R}$. It is also interesting that
the measurement will be independent from the total number of
$\psi(2S)$ decays. With $3\times 10^9$ $\psi(2S)$ events at
BES-III, assuming that no signal events is observed for the
$\psi(2S) \rightarrow U \chi_{c1,2}$ decay, one can measure the
ratio ${\cal R}$ to be $3\times 10^{-7}$ at 90\% confidence level.
For the $U$ boson search, we can reach $\epsilon \sim  10^{-3}$
for in the mass range less than 176 MeV.

 \vspace*{0.3cm}

{\bf 4. The higgs'-strahlung process $J/\psi \rightarrow Uh^\prime$  }\\
\vspace*{0.2cm}

A possible existence of a light Higgs' boson has been discussed in
Ref.~\cite{batell:2009}. At the BES-III, an interesting process is
the higgs'-strahlung $J/\psi \rightarrow U h^\prime$, which can be
looked for if $m_U + m_{h^\prime}< m_{J/\psi}$. There are two
cases to consider: either the Higgs' is heavier than that of the
$U$ boson, or vice versa. These cases will lead to different
experimental signatures at BES-III.

 If the case $m_{h^\prime} < m_U$, the Higgs' is extremely narrow
 and very long-lived~\cite{batell:2009}, so that the signature of
 the process will be $J/\psi \rightarrow U h^\prime \rightarrow
 l^+l^- +$missing energy. There are a few advantages for this kind
 of signal. Firstly, the main background from QED $e^+e^-
 \rightarrow l^+l^- \gamma$ events with the photon lost in the
 calorimeter, is suppressed due to the high detection efficiency
 at BES-III. Secondly, since the missed particle is photon, the
 missing momentum will be equal to the missing energy, one can
 define the observable: $U_{\mbox{miss}} = E_{\mbox{miss}} - |P_{\mbox{miss}}|$, which
 peaks at zero for photon, while is far away from zero for Higgs'
 with non-zero mass. Thirdly, in term of both background rejection
 and detection efficiency, the angular distribution of the signal
 process is proportional to sin$^3(\theta)$, which peaks at
 $\theta = \pi/2$~\cite{kleo2009}. With $10^10$ $J/\psi$ decay
 events at BES-III, the reach of $\epsilon$ could be less than
 $10^{-3}$ in this process.

 For the case $m_{h^\prime} > m_U$, the Higgs' decays almost
 exclusively to two $U$ bosons. For $m_{h^\prime}> 2m_U$, three
 pairs of leptons will have an invariant mass peaked very narrowly
 around the mass of the $U$ boson. The process $J/\psi \rightarrow U
 h^\prime \rightarrow 3(e^+e^-)$ will suffer huge background
 from QED process $e^+e^- \rightarrow 3(e^+ e^-)$ which has the production
 cross-section~\cite{qedphyreport}
\begin{eqnarray}
\sigma_{e^+e^-\rightarrow 3(e^+e^-)} &=& \frac{\alpha^6}{\pi^3
m_e^2}\left( \mbox{log}\frac{s}{m^2_e}\right)^4\\ \nonumber &=&
2.4 \times 10^{12} \, \mbox{fb}. \label{3ee}
\end{eqnarray}
However, as discussed in Ref.~\cite{batell:2009}, the signal
events will have large transverse momentum $p_T$. For the
background case, the background events by contrast will have two
electrons down to the beam pipe, with $\theta< m_e /\sqrt{s}$.
Thus, a small angle cut will reduce the background significantly.
From experimental point of view, we would suggest that $J/\psi
\rightarrow U h^\prime \rightarrow 3(\mu^+\mu^-)$ process will be
important since the QED background in this case should be
negligible as it cannot proceed via the two-photon mechanism, and
the cross-section will decrease as $1/s$. Although there will be a
large number of fake events from $e^+e^- \rightarrow e^+e^-
3(\mu^+\mu^-)$ arising from a two-photon process, they can be
easily removed since $e^+e^-$ pair will be lost down the beam pipe
leading to missing energy. Therefore, the process $J/\psi
\rightarrow U h^\prime \rightarrow 3(\mu^+\mu^-)$ could be
efficiently used as a "discovery mode" for the $U$ boson with huge
$J/\psi$ decay events. In the allowed kinematic region, the reach
of this mode could be $\epsilon \sim 10^{-4}$~\cite{darkslac2009}.

\section{Summary}
\label{sec:sum}

Dark matter experiments suggest new low-energy gauge interactions
beyond the Standard Model. If a dark sector exists, it will
dramatically refresh our understanding of the nature.
 In this Letter, following the work in reference~\cite{wang:2009}, we have investigated the signatures of a hidden
U(1)$_d$ sector at the BES-III experiment, and find that the
BES-III should have an intrinsic sensitivity to the kinetic mixing
parameter $\epsilon$ in the range of $10^{-4}-10^{-3}$.
Especially, the Higgs'-strahlung process will be potential mode to
detect multilepton final states, which has the advantage of
leading novel discovery.

Beginning in mid-2008, the BEPCII/BES-III was operated at
center-of-mass energies corresponding to $\sqrt{s} = 2.0 -4.6 $
GeV. The design luminosity over this energy region  ranges from
$1\times 10^{33}$cm$^{-2}$s$^{-1}$ down to about $0.6 \times
10^{33}$cm$^{-2}$s$^{-1}$~\cite{besiiidetector}, yielding around 5
fb$^{-1}$ each year at $\psi(3770)$ above $D\bar{D}$ threshold and
3 fb$^{-1}$ at $J/\psi$ peak in one year's running with full
luminosity~\cite{besiiidetector}. The BES-III will be able to test
the new interaction in detail, and the $U$ boson and light Higgs
can be looked for in the mass range of a few hundred MeV to GeV
scale.

\section*{Acknowledgments}
The authors would like to thank S.~H.~Zhu and P.~F. Yin for
extensive discussions and suggestion. One of the authors,
H.~B.~Li, would like to thank Professor Michael Peskin for his
warm invitation to visit SLAC for the "Dark force workshop", so
that we had opportunity to discuss with M. Graham and L.~T.~Wang
on the potential for probing the light U-Boson at BES-III. This
work is supported in part by the National Natural Science
Foundation of China under contracts Nos.
10521003,10821063,10979008,10835001, the National Key Basic
Research Program (973 by MOST) under Contract No. 2009CB825200,
Knowledge Innovation Key Project of Chinese Academy of Sciences
under Contract No. KJCX2-YW-N29, the 100 Talents program of CAS,
and the Knowledge Innovation Project of CAS under contract Nos.
U-612(IHEP).




\end{document}